# Renormalization of Bulk Magnetic Electron States at High Binding Energies

A. Hofmann[1], X.Y. Cui[1,2], J. Schäfer[1], S. Meyer[1], P. Höpfner[1], C. Blumenstein[1], M. Paul[1], L. Patthey[2], E. Rotenberg[3], J. Bünemann[4], F. Gebhard[4], T. Ohm[5], W. Weber[5], and R. Claessen[1]

[1] *Physikalisches Institut, Universität Würzburg, 97074 Würzburg, Germany*
[2] *Swiss Light Source, Paul-Scherrer Institut, 5232 Villigen, Switzerland*
[3] *Advanced Light Source, Lawrence Berkeley National Laboratory, Berkeley, California 94720, USA*
[4] *Fachbereich Physik, Philipps-Universität Marburg, 35032 Marburg, Germany*
[5] *Fakultät Physik, Technische Universität Dortmund, 44221 Dortmund, Germany*

The quasiparticle dynamics of electrons in a magnetically ordered state is investigated by high-resolution angle-resolved photoemission of Ni(110) at 10 K. The self-energy is extracted for high binding energies reaching up to 500 meV, using a Gutzwiller calculation as a reference frame for correlated quasiparticles. Significant deviations exist in the 300 meV range, as identified on magnetic bulk bands for the first time. The discrepancy is strikingly well described by a self-energy model assuming interactions with spin excitations. Implications relating to different electron-electron correlation regimes are discussed.

PACS numbers:

The many-body ground state of condensed matter is reflected in its single-particle excitations, which in many cases are significantly modified by coupling to collective modes. Such interactions lead to a pronounced change in the quasiparticle (QP) band dispersion, a so-called *kink*. In metals, the kink from electron-phonon coupling is well-established [1]. Energy-renormalization is also found in the high-$T_c$ cuprates [2]. However, the nature of this feature is not yet completely clarified, albeit of primordial importance to the mechanism of superconductivity. It is being discussed whether the kink is derived from coupling to phonons or to spin fluctuations [3]. Their similar energy scales in the cuprates make it difficult to separate the contributions. Recent experiments on relationships with sample parameters [4, 5] argue for the magnetic coupling model. Closer resemblance to magnetic metals is found in the newly discovered iron-based pnictides [6]. Their parent compounds are true metals with delocalized electrons forming an antiferromagnetic spin density wave. A pairing mechanism based on spin fluctuations has been suggested [7].

Interestingly, a different explanation for kinks in strongly correlated electron systems was suggested recently, which does not require electron-boson coupling. Calculations based on pure electron-electron interaction found two well-separated regimes of QP renormalization [8]. Near the Fermi level, well-defined QPs exist according to Fermi liquid theory. Beyond a characteristic energy scale, the slope of the electronic self-energy changes abruptly, resulting in reduced QP lifetimes and energy renormalization. In the transition between these situations, a dispersion anomaly is expected to emerge [8].

In order to gain access to these many-body interactions, a three-dimensional Fermi liquid in the ferromagnetic state seems a suitable model system. The energy scales for the lattice and spin wave excitations in typical ferromagnets such as Ni differ by approximately an order of magnitude, and hence will affect the QPs at different binding energies [9,10]. Furthermore, it is established that the valence band states are strongly correlated [11,12], which is proven by a concomitant photoemission satellite. This allows to directly adress the interplay of correlation physics and QP formation in the presence of distinct spin excitations.

In this Letter, we present a high-resolution angle-resolved photoelectron spectroscopy (ARPES) study of QP states in Ni(110). Despite the ubiquitous problem of perpendicular momentum broadening, suitable observation conditions can be found for *bulk* bands. An electronic self-energy analysis is carried out by comparison with a Gutzwiller calculation which describes QP renormalization more reliably than state-of-the-art density functional theories [13,14]. The ARPES data show a characteristic high-energy anomaly peaking at ~270 meV. This effect points at substantial interactions between electrons and spin excitations, and calls for refined theoretical descriptions.

The experiments comprise a k-space survey performed at the Advanced Light Source at beamline 7, and high-resolution studies at the Swiss Light Source at the SIS beamline. A clean Ni(110) crystal was prepared by repeated Ar sputtering (1 keV) and annealing (600 °C) *in situ* at a base pressure in the $10^{-11}$ mbar range. Using a Scienta SES-2002 analyzer, total energy resolution amounted to 20 meV, with the sample at T = 10 K. To ensure validity of the free-electron final state approximation, a high photon energy of 100 eV was chosen. The polarization was linear along the [001] crystal direction.

For the purpose of studying QP effects in a wide energy range, the minority spin band at the K point (labeled $\Sigma_2$) was selected, see Fig. 1(a). The ARPES data in Fig. 1(b) are unperturbed by neighboring states for binding energies up to 800 meV. In the $k_\parallel$-plane, the $2^{nd}$ Bril-louin zone (BZ) proved a suitable place for measurement of the $\Sigma_{2\downarrow}$ band, and yielded a higher intensity than in the 1$^{st}$ BZ. Orientation in k-space was provided by a Gutzwiller Fermi surface calculation including $k_\perp$ as in Fig. 1(c).

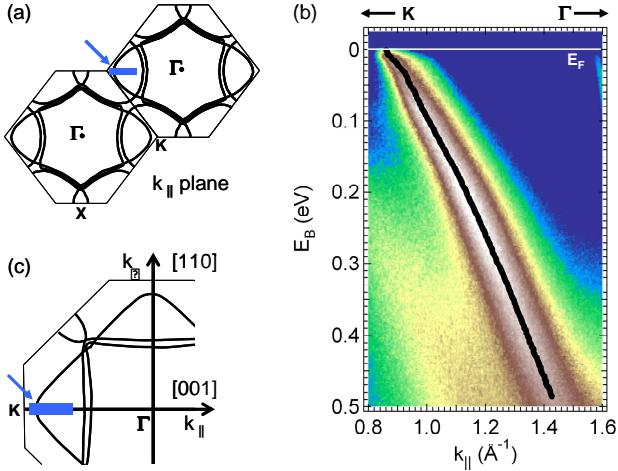

FIG. 1. (Color online) (a) Symmetry plane along $k_\parallel$ in Ni(110) with schematic Fermi surface, showing ARPES scan at K point in 2$^{nd}$ BZ (blue). (b) ARPES band map of $\Sigma_{2\downarrow}$ band along $\Gamma$-K, taken h$\nu$ = 100 eV. Black line indicates MDC peak positions. (c) Gutzwiller Fermi surface cut in $k_\perp$-plane, highlighting curvature perpendicular to the surface.

In photoemission from bulk states, one has to take into account an energy broadening of the spectra, induced by the finite escape depth of the photoelectron. The damping leads to a broadening of the final state, showing a Lorentzian distribution in $k_\perp$ with a full width at half-maximum (FWHM) of $\delta k_\perp = w^{-1}$, where w is the mean free path of the electron. This results in a total energy broadening of the spectra with contributions from the hole state and the photoelectron [15]: $\Gamma_{tot} \approx \Gamma_h + |v_h/v_e|\Gamma_e$, where $v_{e,h} = \partial E_{e,h}/\partial k_\perp$ are the velocities of electrons and holes, respectively.

Hence, the final state might influence the photoemission spectrum significantly, unless the dispersion $v_h$ of the initial state perpendicular to the surface is vanishing. This is fulfilled e.g. for surface states. Yet even for bulk states, the crystal symmetry can be favorably exploited. Selecting the photon energy such that $k_\perp$ lies in a high symmetry plane implies an extremum in the band topology. Thus $v_h$ will vanish, leaving $\Gamma_h$ as the desired QP contribution of interest.

High symmetry planes in $k_\perp$ were determined by variation of photon energy. Still the momentum distribution curves (MDCs) near the Fermi level show a slightly asymmetric line shape. We ascribe this to the $k_\perp$-curvature of the Fermi surface (Fig. 2). Due to the window for $k_\perp$-broadening, neighboring states contribute to the photoemission intensity [16].

A simulation of this effect on states near the Fermi level is shown in Fig. 2(a). It depicts the Gutzwiller Fermi surface curvature at the K point, and assumes MDCs of given width typical for data close to $E_F$ (FWHM 0.05 Å$^{-1}$). Intensity from states surrounding the $\Gamma$-K-line is accounted for by summing up the MDCs weighted by a Lorentzian of FWHM $\delta k_\perp$ = 0.25 Å$^{-1}$. This modeling leads to a weakly asymmetric MDC, Fig. 2(b), whose peak position is virtually unaffected. For ARPES data analysis, the shape can be fitted well in analytic form by an asymmetric Lorentzian [17]. Indeed, it fits the experimental line shape in Fig. 2(c) very well.

In order to extract a MDC width for QP analysis, we use the half-width at half maximum (HWHM) on the narrow side of the peak. This reduced HWHM ignores most of the unwanted $k_\perp$-broadening and gives a realistic measure of the underlying peak width.

For analysis of the many-body interactions, the self-energy $\Sigma$ contains information about the QP properties of the photohole. The real part depicts the mass renormalization of the band dispersion, whereas the imaginary part corresponds to the energy broadening associated with scattering [1]. A reliable method for determining these quantities is from the peak positions and widths of the MDCs. Then, $\text{Re}\,\Sigma(\omega) = \varepsilon(k) - \omega$, where $\varepsilon(k)$ is the bare band dispersion, and $|\text{Im}\,\Sigma(\omega)| = |v(k)|\Delta k$, where $v(k)$ is the slope of the bare band and $\Delta k$ the HWHM of the peak.

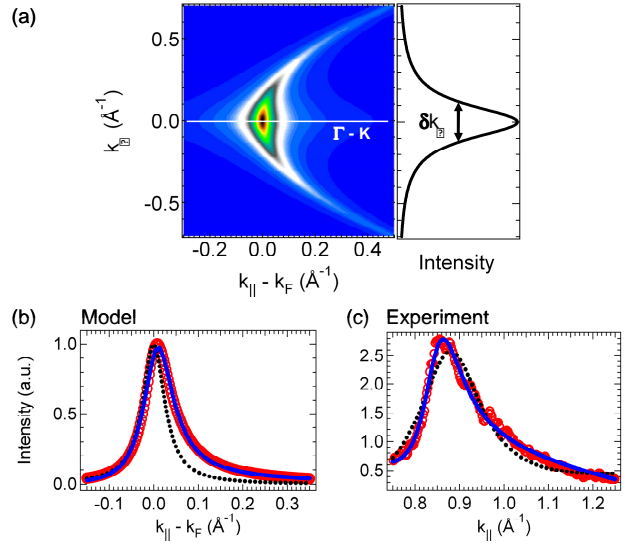

FIG. 2. (Color online) Effect of $k_\perp$-broadening: (a) $\Gamma$-K symmetry line in Gutzwiller Fermi surface, with weighted MDCs (FWHM 0.05 Å$^{-1}$) as intensity plot (for w = 4 Å). (b) Modelled effective MDC from weighted sum of MDCs (red circles). Compared to a symmetric Lorentzian (black dots), the asymmetric Lorentzian is a good analytical fit (blue line). (c) ARPES data along $\Gamma$-K (red circles). The deviation from symmetric behavior (black dots) is well fitted by the asymmetric Lorentzian (blue line).

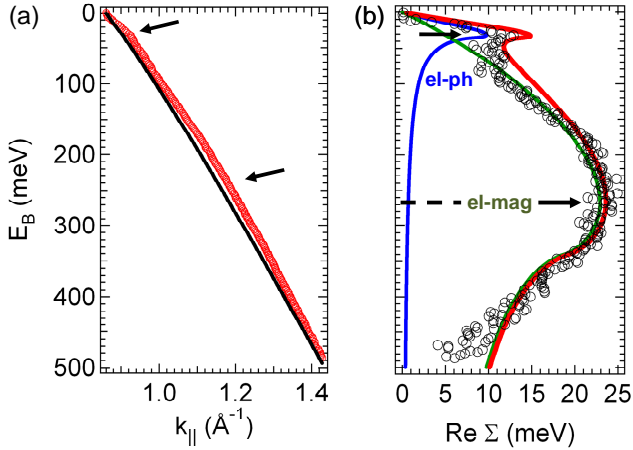

FIG. 3. (Color online) (a) ARPES band data at Ni(110) K-point, represented by fitted MDC peak positions (red), and calculated Gutzwiller dispersion (black). Two energy ranges with deviations are apparent. (b) Extracted real part of the self-energy (black circles). It can be described by adding contributions from electron-phonon (blue) and electron-magnon interaction (green), leading to the red sum curve.

For fitting MDCs from ARPES, the asymmetric Lorentz function was used. For extraction of the self-energy contributions, a Gutzwiller calculation for Ni was used as reference band [18], as plotted in Fig. 3(a). In the Gutzwiller theory, electron-electron correlation effects are accounted for in a much better way than in density functional theory (DFT) calculations. Hence, deviations of the experimental dispersion should result either from coupling of the electrons to bosonic excitations or electron correlation mechanisms not included in the theoretical model.

Comparing experimental data and calculation, an additional mass renormalization can be found around $E_B$ = 30 meV, which corresponds to the kink expected for electron-phonon coupling [19]. The visibility of this feature proves the high quality of the data. It demonstrates that many-body interactions are indeed observable in bulk bands at high symmetry planes.

A second deviation from the calculation is found in the binding energy region from 250 – 300 meV. This behavior is also reflected by the real part of the self-energy in Fig. 3(b), which is showing two maxima and corresponds very well to the structure of the imaginary part. In Fig. 4(a), the raw MDCs with fit curves are displayed. The extracted $Im\Sigma$ in Fig. 4(b) exhibits two steps associated with the real part maxima.

Hence, there is strong indication of a second many-body interaction effect at higher binding energies in both parts of the self-energy. Possible explanations in a ferromagnet with correlated electrons like Ni are the coupling of the electrons to spin-wave excitations or the inherent behavior of correlated electron systems.

The energy scale for magnons in Ni is known from inelastic neutron scattering, which yields values up to 250 meV [10]. However, this method is problematic for scattering at large wave vectors, which makes it difficult to obtain data near the edges of the BZ. Therefore, one expects an even larger energy for the maximum of the spin-wave spectrum. Cut-off energies obtained by various calculations are 360 meV [20], 370 meV [21] and 270 meV [22].

In an attempt to model the observed characteristics of the self-energy, the simple relation known from electron-phonon coupling is used for the electron-boson coupling effects [1,3]: $Im\Sigma(\omega) \propto \lambda \int_0^\omega \rho_{bos}(\omega')d\omega'$, where $\rho_{bos}$ is the boson density of states and $\lambda$ is the coupling constant. From the linear dispersion for bulk phonons and the quadratic one for bulk magnons, up to their respective cutoffs $\omega_0$, it follows $\rho_{ph} \propto \omega^2$ and $\rho_{mag} \propto \omega^{1/2}$, respectively. These results can be used to calculate $Im\Sigma$ contributions from electron-boson interaction with two parameters, the coupling strength $\lambda$ and the cut-off energy $\omega_0$. The corresponding real parts are calculated directly employing the Kramers-Kronig relation. The results for $Re\Sigma$ and $Im\Sigma$ are overlaid in Fig. 3(b) and Fig 4(b), respectively. For a complete model description of the imaginary part, one has to include the electron-electron scattering contribution, for which a quadratic dependence on the binding energy obtained from Fermi liquid theory is assumed. Additional effects such as impurity scattering and experimental resolution can be accounted for by a constant offset. These contributions add up linearly to the full imaginary part.

With this model, the main features of both self-energy parts are very well reproduced, see Fig. 3(b) and 4(b). The parameters in the model are $\omega_{0,ph}$ = 35 meV and $\omega_{0,mag}$ = 340 meV for the cut-off energies. The latter is compatible with the available bulk spin-wave data [10,20,21,22]. It includes the coupling constants $\lambda_{ph}$ = 0.3 and $\lambda_{mag}$ = 0.19, where $(1+\lambda) = m^*/m$ is the mass enhancement. It is important to note that the spectral features and magnitudes for both Re$\Sigma$ and Im$\Sigma$, respectively, are directly connected due to the *Kramers-Kronig relation*. These two self-energy components *simultaneously* follow our experimental data. This is a stringent criterion for the validity of the bosonic excitation scenario and shows that the electron-magnon coupling model is capable of describing the observed features.

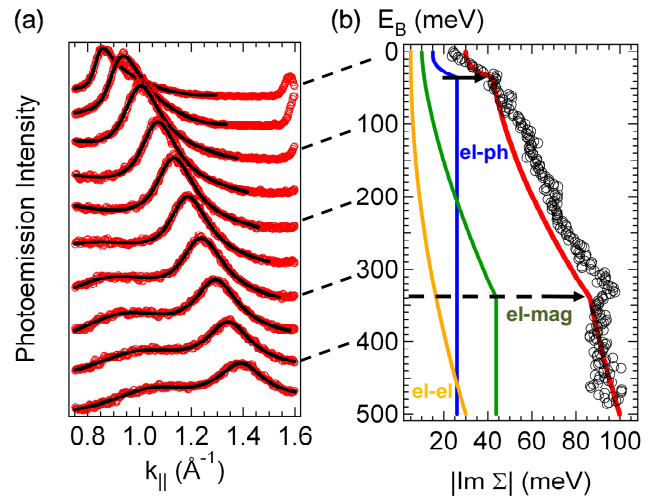

FIG. 4. (Color online) (a) Raw MDC data (red) corresponding to Fig. 3(a), and fitted curves employing the asymmetric peak model (black). (b) Imaginary part of self-energy (black). It shows two shoulders that can be described by electron-phonon (blue) and electron-magnon coupling (green), Kramers-Kronig consistent with $Re\Sigma$ of Fig. 3(b). The sum curve (red) including a small electron-electron contribution (yellow) is a good description of the ARPES data.

This finding and its quantitative analysis significantly extends earlier ARPES observations on surface states of Fe(110) [23]. Indication of electron-magnon interaction was found for *surface* bands, yet at a much lower energy scale. The characteristic energy of 160 meV was subsequently reproduced by spin-polarized electron scattering [24]. Furthermore, our interpretation is in line with quite a number of theoretical works, e.g. on cuprates, in which the emergence of kinks has been attributed to the coupling of QPs and bosonic (antiferromagnetic) spin excitations [25,26,27].

As an alternative approach, it was shown in reference [8] that kinks can also appear solely due to the frequency dependence of the local self-energy within the dynamical mean field theory. The results in [8] allow an estimate of the binding energy at which the kink would be expected in Ni. It is given by the relation $\omega^* = Z_{FL}(\sqrt{2}-1)D$, where $Z_{FL}$ is the low-energy renormalization factor from electronic correlation and D is half the non-renormalized bandwidth. For the case of the $\Sigma_{2\downarrow}$ band, $Z_{FL} \approx v_F^{GT}/v_F^{DFT} \approx 0.78$ is roughly obtained from the Fermi velocities of our Gutzwiller theory and DFT calculations, and D = 1.04 eV from DFT. This coarse approximation yields $\omega^* \sim 330$ meV, which is in the same range as the observed kink energy in Re$\Sigma$ of 270 meV.

However, the kink scenario in [8] was developed for a paramagnetic single-band model in infinite dimensions with Coulomb parameters which gave rise to a strong QP renormalization (Z $\sim$ 0.15). In contrast, Nickel is a ferromagnetic multi-band system with only a moderate renormalization of QP masses. Therefore, it remains open whether or not the approach developed in [8] is of relevance for the kinks observed in Nickel.

This points at the fact that probably the boundary between coupling to a well-defined spin excitation on one hand and coupling between electrons on the other hand might not sharply be drawn. Note that this does not question the existence of well-behaved magnons seen in scattering experiments [10], but rather pertains to the nature of the QP. The interaction model of Byczuk et al. [8] may be understood as a different approach to the same QP phenomenon. In fact, a recent theoretical study [28] argues that this kink formation can be viewed as resulting from emerging *internal* collective excitations.

In conclusion, our experiments show that it is feasible to deduce QP interactions even in three-dimensional solids, using a technique based on high-symmetry planes. The self-energy extracted by a Gutzwiller reference satisfies the Kramers-Kronig citerion. It reveals characteristic structure in the QP dispersion at energies as high as 300 meV and beyond. Its consistency with magnon energies spurs discussion about an accurate theoretical description of the nature of such QPs.

We gratefully acknowledge enlightening discussion with D. Vollhardt and M. Kollar, and support by the EU for travel to the SLS.

[1] see e.g. T. Valla *et al.*, Phys. Rev. Lett. **83**, 2085 (1999).
[2] A. Lanzara *et al.*, Nature **412**. 510 (2001).
[3] A. Damascelli *et al.*, Rev. Mod. Phys. **75**, 473 (2003).
[4] S. V. Borisenko *et al.*, Phys. Rev. Lett. **96**, 067001 (2006).
[5] V. B. Zabolotnyy *et al.*, Phys. Rev. Lett. **96**, 037003 (2006).
[6] Y. Kamahari *et al.*, J. Am. Chem. Soc. **130**, 3296 (2008).
[7] I. I. Mazin *et al.*, Phys. Rev. Lett. **101**, 057003 (2008).
[8] K. Byczuk *et al.*, Nature Phys. **3**, 168 (2007).
[9] A. Dal Corso and S. de Gironcoli, Phys. Rev. B **62**, 273 (2000).
[10] H. A. Mook and D. M. Paul, Phys. Rev. Lett. **54**, 227 (1985).
[11] M. Magnusson *et al.*, Phys.Rev. B **60**, 2436 (1999).
[12] W. Eberhardt and E.W. Plummer, Phys. Rev. B **21**, 3245 (1980).
[13] J. Bünemann *et al.*, Phys. Rev. B **57**, 6896 (1998).
[14] J. Bünemann *et al.*, Phys. Rev. Lett. **101**, 236404 (2008).
[15] J. A. Knapp *et al.* Phys. Rev. B **19**, 4952 (1979).
[16] V. Strocov, J. El. Spec. **130**, 65 (2003).
[17] B. Nemet and L. Kozma, J. Anal. Chem. **355**, 904 (1996).
[18] J. Bünemann *et al.*, Europhys. Lett. **61,** 667 (2003).
[19] M. Higashiguchi *et al.*, Phys. Rev. B **72**, 214438 (2005).
[20] S. V. Halilov *et al.*, Phys. Rev. B **58**, 293 (1998).
[21] R. Brown *et al.*, J. App. Phys. **85**, 4830 (1999).
[22] K. Karlsson and F. Aryasetiawan, Phys. Rev. B **62**, 3006 (2000).
[23] J. Schäfer *et al.*, Phys. Rev. Lett. **92**, 097205 (2004).
[24] W. X. Tang *et al.*, Phys. Rev. Lett. **99**, 087202 (2007).
[25] A. Macridine *et al.*, Phys. Rev. Lett. **99**, 237001 (2007).
[26] F. Tan *et al.*, Phys. Rev. B **76**, 054505 (2007).
[27] Y. Kakehashi and P. Fulde, J. Phys. Soc. Jpn. **74**, 2397 (2005).
[28] C. Raas *et al.*, Phys. Rev. Lett. **102**, 076406 (2009).